\documentclass[sigconf,balance=false]{acmart}
\usepackage{popets}
\setcopyright{popets}
\copyrightyear{YYYY}
\acmYear{YYYY}
\acmVolume{YYYY}
\acmNumber{X}
\acmDOI{XXXXXXX.XXXXXXX}
\acmISBN{}
\acmConference{Proceedings on Privacy Enhancing Technologies}
\usepackage{longtable}
\usepackage{booktabs}
\usepackage{xcolor}
\usepackage{tabularx}
\usepackage{tcolorbox}
\usepackage{boxedminipage}
\usepackage{xparse}
\usepackage{xspace}
\usepackage{multirow}
\usepackage{graphicx}
\usepackage{algorithm}
\usepackage{algpseudocode}
\usepackage{hyperref}
\usepackage{dblfloatfix}
\usepackage{caption}

\newif\iffeedback
\feedbacktrue

\iffeedback

\newcommand{\new}[1]{\textcolor{black}{#1}}
\newcommand{\varun}[1]{\textcolor{red}{Varun: #1}}
\newcommand{\gauthami}[1]{\textcolor{olive}{GY: #1}}
\newcommand{\kevin}[1]{\textcolor{teal}{KY: #1}}
\newcommand{\jj}[1]{\textcolor{blue}{JJ: #1}}
\newcommand{\jt}[1]{\textcolor{orange}{JT: #1}}
\newcommand{\cl}[1]{\textcolor{purple}{CL: #1}}
\newcommand{\by}[1]{\textcolor{violet}{BY: #1}}

\else
\newcommand{\varun}[1]{}
\newcommand{\gauthami}[1]{}
\newcommand{\kevin}[1]{}
\newcommand{\jj}[1]{}
\newcommand{\jt}[1]{}
\newcommand{\cl}[1]{}
\newcommand{\by}[1]{}
\fi

\newenvironment{squishitemize}
{\begin{list}{\textbullet}{%
    \setlength{\itemsep}{0pt}%
    \setlength{\parsep}{0pt}%
    \setlength{\topsep}{0pt}%
    \setlength{\parskip}{0pt} %
    \setlength{\labelwidth}{.5in}%
    \setlength{\labelsep}{0.05in} %
    \setlength{\leftmargin}{.15in} %
    }}
  {\end{list}}

  \newenvironment{squishenumerate}
  {\begin{list}{\arabic{enumi}.}{%
    \usecounter{enumi}%
    \setlength{\itemsep}{0pt}%
    \setlength{\parsep}{0pt}%
    \setlength{\topsep}{0pt}%
    \setlength{\parskip}{0pt}%
    \setlength{\labelwidth}{.5in}%
    \setlength{\labelsep}{0.05in}%
    \setlength{\leftmargin}{.2in}}}
  {\end{list}}

\makeatletter
\long\def\@makecaption#1#2{%
  \vskip\abovecaptionskip
  \footnotesize
  \sbox\@tempboxa{#1.~#2}%
  \ifdim \wd\@tempboxa >\hsize
    #1.~#2\par
  \else
    \hbox to\hsize{\hfil\box\@tempboxa\hfil}%
  \fi
  \vskip\belowcaptionskip}
\makeatother

\usepackage{xcolor}
\usepackage{mdframed}

\newmdenv[
  linewidth=0pt,
  leftline=true,
  rightline=false,
  topline=false,
  bottomline=false,
  linecolor=gray!60,
  linewidth=2pt,
  innerleftmargin=2pt,
  innerrightmargin=0pt,
  innertopmargin=0pt,
  innerbottommargin=0pt,
  skipabove=1pt,
  skipbelow=1pt,
  splittopskip=0pt,
  splitbottomskip=0pt,
  backgroundcolor=white
]{leftquote}

\begin{document}

\title{Understanding How University Guidelines Address Privacy and Security Issues of Generative AI in Academic Settings}

\author{Bei Yi Ng, Jiarui Li, Xinyuan Tong}
\affiliation{%
  \institution{University of Edinburgh}
  \city{}
  \state{}
  \country{}
}

\author{Kevin Ye, Gauthami Yenne, Varun Chandrasekaran}
\affiliation{%
  \institution{University of Illinois Urbana–Champaign}
  \city{}
  \state{}
  \country{}
}

\author{Jingjie Li}
\affiliation{%
  \institution{University of Edinburgh}
  \city{}
  \country{}
}

\begin{abstract}

Generative artificial intelligence (GenAI) is transforming the educational landscape by augmenting learning paradigms. However, state-of-the-art GenAI systems driving this transformation are predominantly developed and controlled by a small number of private companies; there is little clarity about their data retention practices and limited user control over inputs and outputs. In the context of education, end-users lack the awareness of how to safely adopt GenAI in learning. This raises significant concerns, particularly when proprietary or personally identifiable educational information may be shared with external GenAI platforms. In response to these concerns, universities are developing their own usage guidelines and policies to balance innovation with academic integrity, privacy, and security. Our research seeks to understand these emerging guidelines, with a particular focus on the privacy and security implications of integrating GenAI tools into academic environments -- an area that has received little attention to date. We conducted an in-depth qualitative analysis of GenAI-usage guidelines from 43 universities across 12 countries. Our findings reveal several key challenges, including barriers faced by universities in deploying privacy measures and adopting existing security frameworks. These insights lay the groundwork for designing more robust, privacy-aware GenAI guidelines for higher education.

\end{abstract}

\maketitle

\section{Introduction}
\label{sec:intro}

Generative AI (GenAI) is rapidly transforming higher education; students and educators use it for writing, programming, research, and data analysis~\cite{jurenka-arxiv24}. Its versatility in addressing open-ended queries has made it a core tool in academic workflows~\cite{hu2025generative}. As adoption grows, traditional educational technology providers are embedding GenAI into services like writing support and virtual collaboration~\cite{chang2024survey}. In parallel, extensive research is underway to better understand AI-human collaboration in education---ranging from the design of intuitive interfaces to investigations of GenAI's pedagogical impact to the broader shifts it introduces in teaching and learning paradigms~\cite{GenAI2025HigherEd,Quartararo2023GenAIImpact,Cornell2023GenAIReport}.

GenAI systems continually improve by leveraging vast datasets~\cite{liu2024understanding}, including (potential) user interactions~\cite{OpenAI2024DataControls}. In academic settings, GenAI is integrated into tools like AI-enhanced learning management systems and adaptive tutoring applications. However, GenAI providers' (e.g., OpenAI) data practices often lack transparency, raising serious concerns~\cite{Hardinges2024Transparency}. For example, the collapse of AllHere Education---the company behind Los Angeles Unified School District's AI chatbot `Ed'---sparked questions about the fate of sensitive student data it collected~\cite{edsource-url25}. Such incidents highlight the need for stronger oversight in handling educational data. In the U.S., the Family Educational Rights and Privacy Act (FERPA)~\cite{ferpa2011} regulates how student data is accessed and shared, while similar protections exist globally: the EU's General Data Protection Regulation (GDPR)~\cite{GDPR-url25} enforces strict data minimization and consent requirements, and Australia's Privacy Act 1988~\cite{oaic-url25} governs the handling of personal information in schools. These regulatory frameworks compel institutions to implement robust safeguards when adopting GenAI.

While the benefits of GenAI are undeniable---enhancing productivity, creativity, and efficiency---its adoption is accelerating within a ``lemon market,'' where users lack clear visibility into model quality, risks, and trade-offs~\cite{chan2023students}. Students and educators often rely on GenAI tools without a full understanding of their limitations, such as susceptibility to misinformation~\cite{huang2025survey}, bias~\cite{kadaruddin-ijble23}, and over-reliance~\cite{bastani2024generative}. As policy lags behind technological advancement, universities have a vital role to play in shaping responsible GenAI adoption through clear institutional guidelines, safeguards, and pedagogical frameworks~\cite{chan2023students}. Apart from principles of academic integrity, recent studies identify {\em privacy and security} as key motivators in universities' and instructors' policies about GenAI usage~\cite{Jin_Yan2025, McDonald_Johri_Ali_Collier_2025, ali2025-cse25, jiao-nature25}. However, an in-depth understanding of how universities approach and assess these privacy and security risks of GenAI, and in particular how they communicate the trade-offs between potential mitigation strategies and benefits, is still missing. 

Our study aims to bridge this gap. By analyzing GenAI usage guideline documents published online by universities globally, we aim to address the following three research questions:
\begin{squishitemize}
\itemsep0em
\item[\textbf{RQ1.}] How do university guidelines conceptualize and communicate privacy and security risks of GenAI, such as sensitive data breach and technical vulnerabilities?
\item[\textbf{RQ2.}] How do universities evaluate and suggest different measures to mitigate the aforementioned risks?
\item[\textbf{RQ3.}] How are privacy and security positioned among other guiding principles in policy design?
\end{squishitemize}

\new{We conducted qualitative content analysis for university-level GenAI guidelines from 43 universities spanning 12 countries, focusing on how privacy and security considerations integrate with core academic usage and values of GenAI.  Our study reveals in-depth findings corresponding to the privacy and security issues of GenAI in academic settings. We summarize the main findings corresponding to our research questions below:}
\begin{squishitemize}
\itemsep0em

\item [\textbf{F1.}] \new{Universities broadly recognize GenAI-related privacy and security risks, especially those involving sensitive student, research, and institutional data. Their guidelines present partial and evolving threat models, reflecting university priorities: compared with privacy risks from intentional and unintentional misuse by students, staff, and third parties, concrete adversarial behaviors considered as security violations and technical attack surfaces in academic settings are under-specified.}

\item [\textbf{F2.}] \new{Universities mitigate GenAI risks through a combination of technical safeguards, institutional operations, and individual responsibilities, but these measures vary in clarity and consistency. Teaching and learning-focused guidelines often frame privacy and security through academic values and classroom practices, while IT-focused guidelines translate these concerns into more concrete risk assessment, deployment, and monitoring practices. Despite growing reliance on licensed tools and existing security infrastructure, universities still face challenges in adapting safeguards for the expanding ecosystem of GenAI-enabled services.}

\item [\textbf{F3.}] \new{Privacy and security governance is shaped by academic values, institutional policies, and external regulatory or inter-university frameworks, showing evidence of active knowledge transfer. Universities often position privacy and security alongside values such as academic integrity, fairness, and responsible learning, although this values-based framing can leave implementation open to interpretation. Cross-regional and international regulatory influences also shape guideline development, yet the integration of privacy and security with academic values remains uneven across institutions.}

\end{squishitemize}

Based on our findings, we propose several recommendations to improve privacy and security for universities, educators, and students adopting GenAI. These include improving privacy and security assessment frameworks that consider GenAI-specific threats and risks, tailoring technical safeguards to academic contexts, as well as reducing infrastructure and communication disparities for safeguarding GenAI across universities and academic disciplines.

\section{Related Work}
\label{sec:related_work}

\noindent
\textbf{GenAI in education.} GenAI has rapidly gained traction across the higher education sector, permeating both student and faculty practices~\cite{chan2023students,yusuf2024Generativ,francis2025Generative,tillmanns2025mapping,luo2024critical}. Students are leveraging GenAI tools for drafting essays, solving mathematical problems, summarizing readings, generating code, translation, and exam preparation~\cite{Giannakos_Azevedo_Brusilovsky_Cukurova_Dimitriadis_Hernandez-Leo_Järvelä_Mavrikis_Rienties_2024}. Simultaneously, educational professionals are incorporating GenAI into teaching workflows---using it to develop lecture materials, generate quiz banks, provide formative feedback, assist in grading, and personalize instruction through intelligent tutoring systems and adaptive learning platforms~\cite{udeh2025role,qi2025role}. 

Despite this growing adoption, the understanding of GenAI technologies within educational contexts remains fragmented. Many users---students and instructors alike---engage with GenAI tools without a clear grasp of their operational limitations or risks in education. One major concern is the generation of false or misleading content (i.e., hallucinations), which can misinform students or degrade the quality of instruction when integrated into teaching materials~\cite{elsayed2024impact}. This is particularly problematic in activities requiring factual accuracy where hallucinated outputs can introduce conceptual errors or propagate misinformation that is plausible-sounding~\cite{huang2025survey, xie2024survey}.  Another concern is in coding and programming contexts, where GenAI tools may produce code that compiles successfully but fails at runtime, silently introducing logic bugs. In other cases, the code might make use of deprecated libraries, incorrect syntax for a given programming language, or fabricate API calls and functions that do not exist~\cite{liu2024exploring}. 

Instructors may also rely on AI-generated content for assessments or feedback, leading to inconsistent grading, overlooked bias, or inadequate pedagogical alignment~\cite{witsken2025llms}. These risks are amplified in educational settings that inherently demand pluralistic alignment~\cite{baker2021re}---where diverse pedagogical philosophies, disciplinary norms, institutional missions, and cultural values coexist. For instance, the standards for what constitutes a ``good essay'' or ``effective code'' can vary significantly across departments, instructors, and even individual course objectives. 

Finally, there is also the risk of over-reliance, where students defer to GenAI outputs without developing critical thinking or research skills~\cite{bastani2024generative}. Beyond individual use, institutional integration of GenAI into learning management systems or other tools opens the door to system-level vulnerabilities~\cite{greshake2023not}. These risks are exacerbated by the lack of standardized training and clear communication about safe and ethical usage at the institutional level~\cite{zhang2025knowledge, Sheng_2025, Kutty_Chugh_Perera_Neupane_Jha_Li_Gunathilake_Perera_2024}, as well as a dearth of robust yet adaptable AI governance frameworks that can be readily applied in the educational context~\cite{Gajjar_2024,Camacho-Zuñiga_Rodea-Sánchez_López_Zavala_2024, Cao_Fan_Yang_2024}. 

Amidst this landscape, universities are beginning to issue formal guidelines to govern GenAI usage. These institutional policies are becoming essential, not just for setting standards of acceptable use, but for mediating stakeholder differences and fostering a shared understanding of GenAI's role in academic life.

\vspace{1mm}
\noindent
\textbf{Privacy and security risks of GenAI.} Prior research reveals critical privacy and security concerns stemming from emerging GenAI services. The development of GenAI models leads to significant data privacy concerns~\cite{mireshghallah2024trust, IAPP2024PersonalData}, particularly regarding how training data is collected~\cite{yang2024global}. Many GenAI systems are trained on data scraped from the internet without consent, including proprietary or personal materials. In some cases, models can even memorize and regurgitate specific personal or sensitive details from training data, posing risks of inadvertent disclosure of private information~\cite{das-cs25, Williams_2024}.
Furthermore, prior work shows that GenAI introduces exacerbated privacy risks and harms, including inferring/exposing sensitive information and encouraging excessive data collection~\cite{lee-chi24}.
GenAI's data privacy issues are prominent in academic settings too: when students input assignments, questions, or personal reflections into third-party GenAI tools, they may inadvertently expose sensitive academic information—raising potential violations of privacy laws~\cite{parks2017beyond, Wang_Li_Cong_2025}.

GenAI also introduces unique security challenges.
GenAI models are susceptible to jailbreaking, where users manipulate prompts to bypass safety filters and elicit offensive, inappropriate, or dangerous responses~\cite{chu2024comprehensive}.  Furthermore, when GenAI is integrated into tools such as chatbots, it becomes vulnerable to prompt injection attacks~\cite{liu2023prompt}. These attacks involve embedding hidden instructions in input text that can force the model to behave in unintended ways—such as leaking private data, overriding guardrails, generating software with vulnerabilities that are difficult to detect, or even acting on behalf of a user without authorization. 
Furthermore, GenAI is vulnerable to ``poisoning'', since these models require data-intensive training and frequent updates. During model training and update, if a model ingests skewed, offensive, or malicious content injected by a malicious party, it can reproduce those biases in output, leading to discriminatory or harmful responses~\cite{ranjan2024comprehensive}.

\vspace{1mm}
\noindent
\textbf{Privacy and security concerns of educational technologies.} The digitization of education has led to the widespread adoption of educational technologies. From learning management systems (LMS), such as Canvas and Blackboard, to intelligent tutors, to online proctoring tools used in remote assessments. These technologies offer clear benefits in terms of accessibility, scalability, and efficiency~\cite{zhang2024edtechusage}. However, they introduce concerns with respect to security and data privacy, especially in proctored assessment environments and those involving minors~\cite{chanenson2023uncovering, shioji-sp24, Wang_Li_Cong_2025}. There exists a power imbalance between educational institutions and educational technology companies, which often leads institutions to adopt these technologies despite privacy and security concerns and limited resources to investigate them~\cite{kelso2024trust, reidenberg2018privacyined}. Subsequently, students and educators often have little input and control on  what happens to the personal information being collected and stored~\cite{balash2021perceptionsonproctoring, ekambaranathan2021edtechprofit, kwapisz2024shareddata, Wang_Li_Cong_2025, Sheng_2025}. It is common for proctoring tools, for example, to surveil students using key logging, mouse tracking, and microphone and eye tracking ~\cite{burgess2022proctoringvulnerability, shioji-sp24, terpstra2023proctoring}. Similarly, LMS platforms collect data on student behavior and store sensitive and legally protected information such as grades and personal information ~\cite{kwapisz2024shareddata}. In K-12 classrooms, these risks are exacerbated~\cite{kumar2019edtechelementary, zhao2019childrensperception}. For example, in the United States, children under the age of 13 are subject to special legal protections under the Children's Online Privacy Protection Act (COPPA) and Family Educational Rights and Privacy Act (FERPA)~\cite{ftcCOPPA, ferpa2011}. Despite these protections and being less equipped to understand and deal with the consequences of data collection, LMS treat these students the same as others ~\cite{chanenson2023uncovering, kumar2019edtechelementary, maqsood2021tweensonlinerisks, zhong2023k8edtech}. These concerns, compounded by the rapid and expanding adoption of EdTech platforms without thorough evaluation of privacy and security implications, underscore the necessity for scrutiny.

\vspace{1mm}
\noindent \textbf{Research gaps in prior work.}
GenAI is drawing increasing attention in education research. Despite recent studies which identify GenAI privacy and security among other key considerations such as learning enhancement and ethics in the academic usage of GenAI~\cite{Jin_Yan2025, McDonald_Johri_Ali_Collier_2025, ali2025-cse25, jiao-nature25}, they focus on broader pedagogical analyses and consider privacy and security as secondary. Our work contributes novel findings through an in-depth analysis into the technical, operational, and infrastructural challenges to addressing GenAI privacy and security issues through the study of universities' GenAI guidelines -- a gateway to understanding universities' position and implementation of privacy and security measures. Our approach anchors  on how privacy and security concerns are institutionally perceived and systematically managed. This complements existing work that has flagged specific stakeholder concerns, such as that of students and teachers, towards GenAI tools ~\cite{Sheng_2025, Wang_Li_Cong_2025}, as well as emergent literature on how these stakeholders are affected by the ever-changing and vague legislation on implementation ~\cite{Camacho-Zuñiga_Rodea-Sánchez_López_Zavala_2024, García-López_Trujillo-Liñán_2025}. 
 Our study is also distinct from those that study privacy and security issues of other education technologies~\cite{chanenson2023uncovering, shioji-sp24,kumar2019edtechelementary}, as the subject of our study is GenAI, which introduces unique privacy and security risks and challenges in policy enforcement due to its wide and flexible application contexts in education.

\section{Methodology}
\label{sec:method}

{
\scriptsize
\begin{table*}[t]
\centering
\begin{tabular}{p{1.8cm} p{2cm} p{7cm} p{5.2cm}}
\toprule
\textbf{Continent} & \textbf{Country/Region} & \textbf{University} & \textbf{Document Title} \\
\midrule
Africa & South Africa & University of Cape Town (UCT) & \href{https://www.uct.ac.za/sites/default/files/media/documents/uct_ac_za/87/EiRC_GenerativeAI_guideline_Oct2023_final.pdf}{EIRC Generative AI Guidelines} \\
Asia & Hong Kong, PRC & University of Hong Kong (HKU) & \href{https://libguides.lib.hku.hk/AI-literacy/Home}{AI Literacy} \\
Asia & Japan & University of Tokyo & \href{https://utelecon.adm.u-tokyo.ac.jp/en/docs/ai-tools-in-classes-students}{Use of AI Tools in Classes} \\
Asia & Japan & Tokyo Institute of Technology (Tokyo Tech) & \href{https://www.titech.ac.jp/english/student/students/news/2024/069374}{Policy on Use of GenAI in Learning} \\
Asia & Saudi Arabia & King Abdullah University of Science and Technology (KAUST) & \href{https://researchcompliance.kaust.edu.sa/researchintegrity/guidelines/Joint%20statement_11April2023.pdf}{AI Guidelines on Research} \\
Asia & Singapore & Nanyang Technological University (NTU) & \href{https://www.ntu.edu.sg/education/inspire/teaching-learning-assessment-with-genai/assessment/policies-guidelines}{Teaching Learning Assessment with GenAI} \\
Asia & Singapore & Nanyang Technological University (NTU) & \href{https://www.ntu.edu.sg/research/resources/use-of-gai-in-research}{Use of GenAI in Research} \\
Asia & Singapore & National University of Singapore (NUS) & \href{https://libguides.nus.edu.sg/new2nus/acadintegrity}{NUS Academic Integrity} \\
Europe & Germany & Technical University of Munich (TUM) & \href{https://tumthinktank.de/event/usage-guidelines-for-generative-ai-tum/}{Usage Guidelines for GenAI $@$ TUM} \\
Europe & Netherlands & Delft University of Technology (TU Delft) & \href{https://www.tudelft.nl/teaching-support/didactics/assess/guidelines/ai-chatbots-in-unsupervised-assessment}{AI Chatbots in Unsupervised Assessment} \\
Europe & Netherlands & University of Amsterdam (UvA) & \href{https://www.uva.nl/en/about-the-uva/about-the-university/ai/ai-policy/ai-policy.html}{UvA Policy on AI} \\
Europe & Switzerland & École polytechnique fédérale de Lausanne (EPFL) & \href{https://www.epfl.ch/about/vice-presidencies/vice-presidency-for-academic-affairs-vpa/tips-for-the-use-of-generative-ai-in-research-and-education/}{Tips for the Use of GenAI} \\
Europe & Switzerland & Eidgenössische Technische Hochschule Zürich (ETH Zurich) & \href{https://ethz.ch/content/dam/ethz/main/eth-zurich/education/ai_in_education/Generative%20AI%20in%20Teaching%20and%20Learning%20-%20Guidelines%20ETH.pdf}{GenAI in Teaching \& Learning} \\
Europe & United Kingdom & Imperial College London (IC) & \href{https://www.imperial.ac.uk/admin-services/library/learning-support/generative-ai-guidance/}{GenAI Guidance} \\
Europe & United Kingdom & King's College London (KCL) & \href{https://www.kcl.ac.uk/about/strategy/learning-and-teaching/ai-guidance/student-guidance}{AI Guidance on Learning and Teaching} \\
Europe & United Kingdom & London School of Economics and Political Science (LSE) & \href{https://info.lse.ac.uk/staff/divisions/Eden-Centre/Artificial-Intelligence-Education-and-Assessment/School-guidance}{AI Education and Assessment} \\
Europe & United Kingdom & University of Edinburgh (UoE) & \href{https://www.ed.ac.uk/bayes/ai-guidance-for-staff-and-students/ai-guidance-for-students}{AI Guidance for Students} \\
Europe & United Kingdom & University of Manchester (UoM) & \href{https://manchester-uk.libanswers.com/teaching-and-learning/faq/264824}{Teaching and Learning FAQ} \\
Europe & United Kingdom & University College London (UCL) & \href{https://www.ucl.ac.uk/teaching-learning/publications/2023/sep/using-generative-ai-genai-learning-and-teaching}{Using GenAI in learning and teaching} \\
Europe & United Kingdom & University of Bristol  & \href{https://www.bristol.ac.uk/bilt/sharing-practice/guides/guidance-on-ai/}{University Guidance on GenAI in Education} \\
Europe & United Kingdom & University of Cambridge  & \href{https://blendedlearning.cam.ac.uk/guidance-support/ai-and-education/using-generative-ai}{Using Generative AI} \\
Europe & United Kingdom & University of Oxford  & \href{https://communications.admin.ox.ac.uk/communications-resources/ai-guidance#collapse4654526}{Guidelines on the Use of GenAI} \\
North America & Canada & McGill University  & \href{https://www.mcgill.ca/it/article/microsoft-copilot-and-general-guidelines-using-generative-ai-tools}{General Guidelines for Using GenAI Tools
}\\
North America & Canada & University of British Columbia (UBC) & \href{https://academicintegrity.ubc.ca/genai/}{GenAI tools in teaching and learning} \\
North America & Canada & University of Toronto (UofT) & \href{https://www.viceprovostundergrad.utoronto.ca/16072-2/teaching-initiatives/generative-artificial-intelligence/}{GenAI in the Classroom: FAQ's} \\
North America & United States & California Institute of Technology (Caltech)& \href{https://ctlo.CalTech.edu/universityteaching/resources/resources-for-teaching-in-the-age-of-ai}{Resources for Teaching in the Age of AI} \\
North America & United States & Carnegie Mellon University (CMU) & \href{https://www.cmu.edu/computing/services/ai/index.html}{Generative Artificial Intelligence} \\
North America & United States & Columbia University  & \href{https://provost.columbia.edu/content/office-senior-vice-provost/ai-policy}{GenAI Policy} \\
North America & United States & Cornell University  & \href{https://teaching.cornell.edu/generative-artificial-intelligence/cu-committee-report-generative-artificial-intelligence-education}{GenAI for Education and Pedagogy} \\
North America & United States & Duke University  & \href{https://lile.duke.edu/ai-and-teaching-at-duke-2/artificial-intelligence-policies-in-syllabi-guidelines-and-considerations/}{AI Policies: Guidelines and Considerations} \\
North America & United States & Harvard University  & \href{https://huit.harvard.edu/ai/guidelines#:~:text=Protect%20confidential%20data,-You%20should%20not&text=Level%202%20and%20above%20confidential,Security%20and%20Data%20Privacy%20office.}{Generative AI Guidelines} \\
North America & United States & Massachusetts Institute of Technology (MIT) & \href{https://ist.mit.edu/ai-guidance}{Guidance for Use of GenAI Tools} \\
North America & United States & Northwestern University (NU)& \href{https://www.it.northwestern.edu/about/policies/guidance-on-the-use-of-generative-ai.html}{Northwestern Guidance on the Use of GenAI} \\
North America & United States & University of California, Berkeley (UC Berkeley) & \href{https://oercs.berkeley.edu/privacy/privacy-resources/appropriate-use-generative-ai-tools}{Appropriate Use of GenAI Tools} \\
North America & United States & University of California, Los Angeles (UCLA) & \href{https://genai.ucla.edu/guiding-principles-responsible-use}{Guiding Principles for Responsible Use} \\
North America & United States & University of Chicago  & \href{https://its.uchicago.edu/generative-ai-guidance/#:~:text=Generative%20AI%20systems%2C%20applications%2C%20and,and%20privacy%20of%20University%20data.}{GenAI Guidance} \\
North America & United States & University of Illinois Urbana-Champaign (UIUC)   & \href{https://www.vpaa.uillinois.edu/digital_risk_management/generative_ai/principles}{Univ. of Illinois System GenAI Principles} \\
North America & United States & University of Michigan-Ann Arbor (UM-Ann Arbor) & \href{https://genai.umich.edu/guidance/students}{U-M Guidance for Students} \\
North America & United States & University of Pennsylvania (UPenn) & \href{https://www.isc.upenn.edu/security/AI-guidance}{AI Guidance} \\
Oceania & Australia & Monash University  & \href{https://www.monash.edu/learning-teaching/TeachHQ/Teaching-practices/artificial-intelligence/old-version/policy-and-practice-guidance-around-acceptable-and-responsible-use-of-ai-technologies2}{Acceptable and Responsible Use of AI Tech.} \\
Oceania & Australia & Australian National University (ANU) & \href{https://libguides.anu.edu.au/generative-ai}{AI including GenAI} \\
Oceania & Australia & University of Melbourne  & \href{https://www.unimelb.edu.au/generative-ai-taskforce/university-of-melbourne-ai-principles}{University of Melbourne AI Principles} \\
Oceania & Australia & University of New South Wales (UNSW)& \href{https://www.student.unsw.edu.au/skills/ai}{ChatGPT \& GenAI at UNSW} \\
Oceania & Australia & University of Queensland (UQ)& \href{https://itali.uq.edu.au/teaching-guidance/teaching-learning-and-assessment-generative-ai?p=7#7}{AI Teacher Hub } \\
\bottomrule
\end{tabular}
\caption{The list of official GenAI documents we analyzed. These documents are published online by universities worldwide.}
\label{tab:policies}
\end{table*}
}

\new{To answer our research questions in \S~\ref{sec:intro},} we conducted a qualitative analysis of university-level GenAI policy documents (or guidelines). Our approach allows us to characterize the early governance landscape, identify emerging privacy and security measures and risk framings. We then examine how institutions are developing guidance to manage vulnerabilities associated with GenAI adoption. 
We employed a structured qualitative coding and analysis process, enabling us to surface cross-institutional patterns and evolving requirements in privacy and security-aware GenAI integration from a globally representative sample of university guidelines.

\vspace{1mm}
\noindent{\bf Data collection and sampling.} Our data collection focused on publicly accessible GenAI guideline documents published by universities globally. 
We conducted this effort between June and December 2024, and we selected universities from the ``2024 QS World University Rankings''~\cite{qs-2024}, which has been well-established in the methodology of cross-comparative policy studies~\cite{becker-cse19,weichert-cse25} for its global recognition and comprehensive evaluation of teaching, research, and international outlook. 

\new{Our initial exploration, during which we explored availability of GenAI policies in universities by region, discovered that the institutions recognized in the top universities globally also tend to have more accessible and comprehensive GenAI policy documents, which became our focus, aligning with prior findings that systematic adoption of GenAI remains experimental across much of higher education~\cite{jiao-nature25}.
Though starting from the highest ranked institutions, our sampling aimed to achieve institution and region diversity within available data.
In this study, we focused on and searched for \textit{university-level} guidelines available on university websites, rather than department-specific documents, as these university-level guidelines represent the \textit{overarching} principles of one university, contextualizing privacy and security concerns within educational practices, that directs specific adoption and implementation at department levels.}
We manually traversed university websites, including subdomains on Academic Integrity subsuming GenAI guidelines. 
To assist our search, we constructed a set of search queries to obtain additional results from the search engine, incorporating a combination of representative terms with university names, including \textit{``generative AI,''} \textit{``AI,''} \textit{``guidance,''} \textit{``guideline,''} \textit{``policy,''} \textit{``teaching''} and \textit{``learning,''} that reflect the naming conventions used by institutions to ensure accessibility of their policies. 
 Our search logic includes the above terms: \texttt{(``University name'') \& (``Generative AI'' | ``AI'') \& (``Policy'' | ``Guidance'' | ``Guideline'') \& (``Learning'' | ``Teaching'' | ``'')}.

\new{In addition, our scope of sampling and analysis is limited to GenAI guidelines that are publicly available in English. 
We developed the codebook in stages using universities ranked in the top 100 of the QS World University Rankings, supplemented by two highly ranked regional universities, UCT and KAUST. To reduce overrepresentation of US and UK institutions, we randomized the university order and coded them in batches, refining the codebook after each batch. We stopped sampling once our qualitative coding and analysis reached data saturation~\cite{saunders-qq18}; the coding process is described soon after.}
 
{Our sampling process stopped after code saturation, which we monitored along our analysis, at 43 universities; some universities in the QS ranking are not analyzed due to lack of publicly available guidelines.} We show the list of universities and guidelines analyzed in Table~\ref{tab:policies}.
We only include documents whose overall topic is GenAI guidance in contexts of teaching and learning, excluding those where GenAI has partial focus, e.g., President Remark of Kyoto University~\cite{kyotouFY2024Undergraduate}. 

\new{Following the data collection and analysis of university-level guidelines focusing on academic integration, particularly teaching and learning, we collected additional GenAI guidelines provided by the university IT departments, including information security,  for further analysis and cross-comparison. This additional data source allows us to examine if and how universities provide more dedicated technical resources and support to govern the security \& privacy (S\&P) risks. We adopted the same web search and website traversal approach to collect these documents. 
Similarly, the search logic to assist our manual website traversal for this task includes these terms  \texttt{(``University name'') \& (``Generative AI'' | ``AI'') \& (``Policy'' | ``Guidance'' | ``Guideline'')  \& (``Information security'' | ``Acceptable use'' | ``Approved tools'')}.
The above terms are chosen as they correspond to the core concept of privacy and security related to GenAI.
We initially experimented with more search terms that include \textit{``privacy policy''}, \textit{``data protection''}, and \textit{``student data privacy.''} 
However, such terms inconsistently returned website privacy policies, cookie notices, and general student privacy notices rather than  governance documents specific to GenAI issues, and were therefore not included eventually.
As a result, we included additional IT-specific policies for 19 universities to cross-compare with their university-level, teaching and learning-focused guidelines, shown in Table~\ref{tab:it_policies} in Appendix~\ref{ss:appendix-1}.}

\vspace{1mm}
\noindent{\bf Data preparation.} 
Universities present their guidelines as web pages.
To code and analyze these guideline documents efficiently and consistently, we first reformatted the page content using an \texttt{HTML} sanitizer~\cite{pypiClientChallenge}.
We preserved the section headers and paragraphs 
that are relevant to the guideline content. This approach allowed us to segment guideline documents into coherent paragraphs and sections that are suitable for coding and thematic analysis. We manually verified and curated the segmentation when the sections are not well wrapped by \texttt{HTML} headers. Our analysis focused on text content, which presents the most meaningful information in a guideline webpage. Thus, we did not include visual aids.

\vspace{1mm}
\noindent{\bf Qualitative coding and analysis.}
Our analysis started with open coding~\cite{clarke-jpp17}.
Three researchers independently coded guideline documents from 15 universities, looking for guideline content and concepts that are relevant to our research questions.
We first comprehensively captured fine-grained concepts, allowing codes to develop with the initial codebook.
Following this process, all research team members engaged in a series of structured discussions to review, compare, and refine the codes generated individually.
This contributed to our initial codebook. 
We further leveraged guidelines from three universities to improve the initial codebook and resolve discrepancies, e.g., when guideline framings were ambiguous, we resolved them by refining definitions and merging overlapping codes.
Two researchers later applied this codebook to independently code 22 more guidelines.
We used the 22 coded guidelines here to assess the reliability of our codebook.
We computed Cohen's kappa (\textit{$\kappa$}), a standard metric of inter-annotator agreement for these guidelines \new{coded by two coders. 
Our coding attained \textit{$\kappa$} = 0.65, which indicates a substantial agreement~\cite{mchugh-bioch12}. 
Then we continued coding and confirmed saturation with 46 guideline documents in total.
The lead coder also applied the codebook developed to code and analyze the additional IT-focused GenAI guidelines, further confirming the coverage of the codebook.}

\new{Throughout our study, our research team engaged in regular meetings to resolve disagreement in coding and perform thematic analysis. 
We applied codebook thematic analysis~\cite{braun2021can, clarke-jpp17}, which lies between codebook reliability and reflexive thematic analysis. As such, after reaching code saturation, our team continued analysis to develop themes by identifying patterns from coded data. Note that the codebook served as an analytic framework to underpin later thematic analysis, and \textit{$\kappa$} is treated as a pragmatic quality check rather than a claim of full analytic validity. }
Our research teams are formed with members with diverse expertise, including computer security and privacy, education, human computer interaction, and artificial intelligence.
Our expertise enables our analysis to capture codes and themes from comprehensive and complementary perspectives that are essential to our research questions. \S~\ref{sec:findings} discusses categories of codes and themes discovered. We make our codebook available through our project repository with access to documents.~\footnote{\url{https://osf.io/3ezfw/overview?view_only=3f8cad97c179473eb0ccbfa98595a350}}

\section{Study Limitations}
\label{sec:limitations}

We recognize several limitations in our study. 
Our sampling uses the QS World University Rankings, following prior research~\cite{Jin_Yan2025}. However, we acknowledge that the distribution of institutions across the world is skewed due to regional disparities in educational development and internationalization. The top-ranked universities happen to be concentrated in {North America, Europe, East Asia, and Australia}. 
Despite having considered regional diversity in sampling and actively looked for available content in underrepresented regions including Africa and the Middle East, we ended up including more universities where English is a major working language. This is also because we focused on and searched for GenAI guidelines written in English. 
This also ultimately resulted in exclusion of universities from South America.
\new{Furthermore, our sample presents perspectives of globally ranked and research-intensive universities, also early GenAI adopters, rather than the whole sector of higher education, which therefore may not be representative for less-resourced institutions. 
We also acknowledge that the GenAI guidelines we study present the public, overarching position of a university; the actual implementation may differ, and the policies are subject to evolve.} Recognizing the fast-evolving landscape of GenAI development and integration, we aim to provide in-depth qualitative results, rather than quantitative and statistical results to identify challenges faced by the universities in safeguarding GenAI usage. 
Consistent with prior publications~\cite{Jin_Yan2025, McDonald_Johri_Ali_Collier_2025, Wang_Dang_Wu_Mac_2024}, we opt to conduct an in-depth qualitative analysis of university-level guidelines, as it provides an authentic view of the AI governance landscape that reflects links to state-level and national frameworks, which are limited in course/department-level rules. Nevertheless, this approach is limited in uncovering specific privacy and security challenges in curriculum design when adopting GenAI.

\new{ 
We argue that  global influences of sampled universities on education and GenAI adoption may inform less-resourced institutions, and some challenges, e.g., technical and operational barriers, faced by these universities may be amplified for low-resourced institutions.
Nevertheless, we encourage future work to investigate the challenges faced by less-resourced institutions in the real world and to explore other data sources and methods that offer complementary insights, e.g., from individual perspectives of students and staff, while also conducting longitudinal measurements and comparisons of GenAI guideline updates beyond our study's cut-off date, given the dynamic and ever-evolving nature of policy development. 
Furthermore, large-scale studies can consider GenAI guidelines available in other languages beyond English, as well as the longitudinal factors in institutional policy development.}

\begin{figure*}[t]
    \centering
    \includegraphics[width=0.85\textwidth]{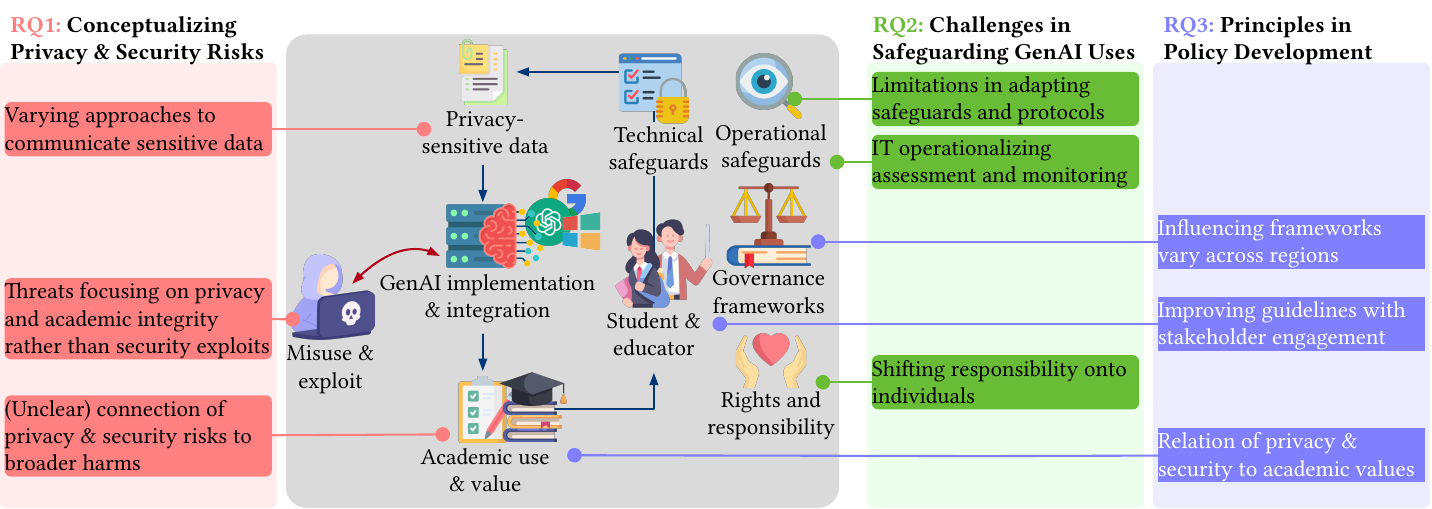}
    \caption{\new{An overview of the findings and themes that address our 3 research questions. We highlight how themes are anchored on 9 categories of codes (within the grey box).} \vspace{-4mm}}
    \label{fig:finding overview}
\end{figure*}
\section{Findings}
\label{sec:findings}

Figure~\ref{fig:finding overview} provides a holistic overview of the observed themes from our analysis, as well as how they answer each research question. These themes highlight the privacy and security challenges from different perspectives in the university integration of GenAI services, and how stakeholders interact with them.

\noindent{\bf Roadmap.} 
\S~\ref{ss:rq1} discusses how guidelines conceptualize GenAI privacy and security risks (\textbf{RQ1}), including sensitive data, technical threats, and connections with broader academic harms.
Recognizing the privacy and security risks, \S~\ref{ss:rq2} elaborates on how universities assess different privacy and security measures (\textbf{RQ2}) from institutional levels to individual levels, specifically challenges in these measures. Finally,
\S~\ref{ss:rq3} explains how privacy and security, as guiding principles, are considered in universities' policy and guideline development (\textbf{RQ3}), along with overarching regulatory frameworks and the tensions with broader academic values.

\subsection{Conceptualizing Privacy and Security Risks}
\label{ss:rq1}

Our qualitative coding discovered three high-level categories of university guidelines' conceptualization of privacy and security risks of GenAI:

\begin{squishenumerate}
\item \textbf{Privacy-sensitive data practices}, which include varying personal or sensitive data types, data handling practices (data collection, third-party sharing, retention, unauthorized sharing of training data etc.), and breaches.
\item \textbf{Exploit and misuse} of GenAI services, including harmful and inaccurate content (misinformation and fraudulent content),  vulnerabilities and bugs (e.g., jailbreaking), and students' over-reliance on GenAI.
\item \textbf{Technical limitations} of GenAI models that can be exploited, such as non-transparency of training, limited reasoning capability  and domain-specificity, and service reliability.
\end{squishenumerate}

\new{Across these categories, university guidelines also implicitly define threat models surrounding the adoption of GenAI
in academic settings. 
Below we further discuss the main (sub)themes observed surrounding elements in the academic threat model, which include (1) protected assets, in particular types of sensitive student and research data, (2) threat actors, especially third-party GenAI vendors and  irresponsible users, and their associated attack vectors, and the key (3) affected stakeholders and systems, particularly the learning processes and outcomes. While universities rarely use formal threat-modeling
terminology, these dimensions appear throughout the guidelines
and shape how privacy and security risks are communicated and
mitigated.}

\subsubsection{\new{Guidelines vary in the approaches to communicate specific sensitive data types to protect.}} \label{ss:rq1-1} Our analysis reveals that while university guidelines broadly recognize the need to regulate handling sensitive data with GenAI tools, the approaches different universities (and guidelines) took to elaborate on the scope of sensitive data vary. Some university guidelines adopt an ``{example-driven}'' approach. For example, Caltech's guideline lists specific examples of sensitive data types including

\begin{leftquote}
    \textit{``...any intellectual property or unpublished research data, export-controlled data, and other sensitive HR, business, or administrative data...''} (Caltech)
\end{leftquote}

However, the scope of sensitive data types discussed can be vague in other guidelines, especially when multiple terminologies related to sensitive data, such as \textit{proprietary information, confidential data, and restricted information}, are used concurrently. These terminologies can be mentioned in the same guideline without further elaboration despite signposting to other internal and/or external policy documents and frameworks. For example, the GenAI guideline of NTU mentions that

\begin{leftquote}
    \textit{``Any confidential or sensitive information, and/or personal data are not to be uploaded to any GAI software, system, or platform...''} (NTU)
\end{leftquote}

while pointing to two other data policies: Singapore's Personal Data Protection Act (PDPA)~\cite{pdpa-url25} and NTU's Data Governance Policy. 
Similarly, MIT prohibits the use of data   \textit{``classified as Medium Risk or High Risk''} using the university's internal data classification framework with AI tools that do not have a license agreement; \S~\ref{ss:rq3} will further discuss how university guidelines leverage existing policy and regulatory frameworks to articulate their own.

Despite the inconsistent terminology and data classifications adopted by universities, university guidelines (such as those from UPenn) broadly highlight the importance of data protection for proprietary and copyright-protected information, which echoes the elevated concerns for GenAI to infringe IP rights~\cite{smits-lai22}:

\begin{leftquote}
    \textit{``Avoid uploading confidential and/or proprietary information to AI platforms prior to seeking patent or copyright protection, as doing so could jeopardize IP rights''} (UPenn)
\end{leftquote}

\subsubsection{\new{ Recognized threat vectors focus on privacy and academic integrity rather than security exploits}}\label{ss:rq1-2}
\new{We find that the threat vectors guidelines recognize are centered on the intentional and unintentional GenAI misuse by service providers and academic users.}

\new{\textit{\textbf{Privacy exposure to third-party service providers passively and actively.}} University guidelines frame third-party providers as a key potential threat actor capable of misusing protected academic and personal data, especially when students and staff unintentionally expose such data.
For example, UBC's guidelines warn that \textit{``anything that is entered into the [GenAI] tool may be inappropriately exposed to third parties''.} 
Universities may explicitly name tools from specific vendors, such as Google Gemini and Microsoft CoPilot, with the associated risk levels based on their licensing and privacy assessment outcomes, which \S~\ref{ss:rq2} will  further elaborate.
}

\new{Moreover, universities are mindful that GenAI service providers may actively infringe users' privacy, as part of the development and technical integration of their GenAI services currently, e.g., \textit{``the risk of exposing personal, confidential and university sensitive information''} (McGill University) for enabling \textit{``internet searching''} (University of Tokyo), frequent model training that infringes \textit{``privacy and intellectual property of information''} (TU Delft), as well as sensitive data required for \textit{``sign[ing] up for an account''} (NUS). }

\new{\textbf{\textit{Concern over students and staff centered on misusing GenAI output against academic integrity.}} On the other hand, guidelines describe the potential of malicious academic use of GenAI by students and staff:}

\begin{leftquote}
\textit{``...Potential for malicious actors to use the tools to facilitate irresponsible research practices linked to fabrication, falsification and plagiarism and to make obfuscation easier...''} (UCT) 
\end{leftquote}

\new{In general, the security threats on the  GenAI model itself are less discussed and less contextualized.  
Compared to the privacy issues of \textit{personal or sensitive information}, which are discussed in 30 universities' guidelines, just 8 mentions potential for \textit{malicious use} with  4 universities discussing \textit{vulnerabilities, bugs and exploits}, including \textit{jailbreaking}.
Among the security issues identified, some university guidelines discuss the secondary security risks due to the malicious use of GenAI-generated content, including those related to harmful textual data and software code. For example, UoE is concerned that GenAI may }

\begin{leftquote}

\textit{``...Create code that has security flaws, bugs and use illegal libraries or calls - or infringe copyrights...''} (UoE) 
\end{leftquote}

Similarly, CMU mentions GenAI's security risks to \textit{``create more convincing phishing emails''} when used by malicious actors as a motivating background beyond the educational context.

\new{\textbf{\textit{User vulnerabilities due to lack of technical understanding of GenAI.}} In addition, some university guidelines are also aware that the technical limitations of GenAI may be challenging for the audience to understand, limiting their ability to safely use the service and leading to privacy or security breaches}. Therefore, these universities leverage analogies and comparison with other technologies that share similar features or concerns to improve reader understanding. For example, NUS reminds the readers of the unreliability of GenAI service providers who \textit{``can sometimes go down without warning''} and compares GenAI with the old-school calculator when discussing its limited scope in use-cases recommended by the guideline.

\subsubsection{\new{Guidelines connect privacy and security risks to broader harms in education which can be unclear.}}\label{ss:rq1-3}
\new{More than the integrated learning management and administrative systems that are affected by privacy breaches, or broader IT infrastructure that is in danger due to security exploits produced by GenAI, universities consistently consider negative impacts on learning processes and outcomes intertwined with privacy and security issues.}

\new{\textit{\textbf{Privacy and security impacts positioned among ethical risks.}} We observe that university guidelines often discuss privacy and security risks within the broader ethical consequences of GenAI that negatively affect students. For example, UIUC names privacy breaches along other harms such as discrimination:}

\begin{leftquote}
    \textit{``...establish procedures for remediations, recourse, or redress in case of unintended consequences, discrimination, or privacy breaches...''} (UIUC)
\end{leftquote}

Many negative consequences which universities are aware of, e.g., misinformation, are unintended due to technical limitations such as hallucination and biased training datasets:

\begin{leftquote}
\textit{``It is also possible for generative AI models to “hallucinate,” or otherwise use false information in their responses''} (NU)
\end{leftquote}

\new{\textit{\textbf{Negative learning outcomes reinforced by adversarial use.}}} Nevertheless, we do observe that several university guidelines realize how ``malicious'' uses of GenAI and its other negative impacts may reinforce each other in students' learning. For instance, Cornell University expresses its  \textit{``overarching concern''} that how students' over-reliance on GenAI might motivate their potential use of ``jailbreaking'' techniques, which is a major security threat in circumventing GenAI guardrails~\cite{wei-neurips23}. This in turn results in students' lack of confidence and ability \textit{``to master needed knowledge or skills.''} More specifically, ``jailbreaking'' could help students in accessing prohibited content in learning:

\begin{leftquote}
\textit{``...Jailbreaking has been used to circumvent prohibitions on certain topics or behaviors such as advocating violence; however, it could also easily be used to obtain answers to homework assignments....''} (Cornell University) 
\end{leftquote}

However, guidelines in general seem less specific on how privacy and security risks may connect to other safety concerns of GenAI.

\subsection{Challenges in Safeguarding GenAI Uses}
\label{ss:rq2}
We observe that university guidelines discuss privacy and security measures for GenAI under the following categories:
\begin{squishenumerate}
    \item \textbf{Technical safeguards}, including AI detection tools, data confidentiality and anonymization, and others such as permission management and access control.
    \item \textbf{Individual rights and responsibility}, such as informed consent, privacy rights and options (e.g., opt-out and data deletion), acknowledging use of AI, and choosing proper alternatives including specific GenAI services to adopt or pedagogical redesign to accommodate GenAI uses. 
    \item \textbf{Operational safeguards}, including security review and licensing, disciplinary actions to ensure compliance of individual conduct and punish misconduct, as well as stakeholder communication and education.
\end{squishenumerate}

This subsection discusses in-depth how university guidelines evaluate the efficacy and employ these measures.

\subsubsection{\new{Guidelines reflect limitations of adapting technical safeguards and existing security protocols.}}\label{ss:rq2-1}
\new{Below discusses the technical reliability concerns and operational complexity to adapt GenAI safeguards we find in the guidelines.}

\new{\textit{\textbf{Reliability concerns of technical safeguards.}}}
Our study reveals that universities are not over-optimistic about the efficacy of AI detection tools when detecting uses of GenAI. Universities such as the University of Cambridge are aware of the limitations of these tools, and do not offer a standard detection tool as 

\begin{leftquote}
\textit{``Detection of AI-generated content is possible, but difficult and not wholly reliable''} (University of Cambridge)
\end{leftquote}

Moreover, some universities are concerned about the instructors' use of AI detection tools due to additional privacy concerns about how students' personal data are handled by these tools.

\begin{leftquote}
\textit{``Instructors should not use these tools to evaluate any student work that contains the name of the student or any other personal information of the student or third parties.''} (UBC)
\end{leftquote}

In contrast to the discouragement of unreliable AI detectors, we observed that universities leverage existing technical infrastructures that are deemed more reliable to secure GenAI when integrating it into their IT systems, e.g., by account verification:

\begin{leftquote}
\textit{``Copilot is an enterprise version of an AI-powered chatbot and search engine which better protects the privacy and security of users (when users are signed into their UofT account).''} (UofT)
\end{leftquote}

Likewise, Caltech's guideline describes their access control in place to \textit{``disable or limit access to AI companion tools''} through their enterprise control. Universities also demand the GenAI service providers to implement GenAI safeguards by design:

    \begin{leftquote}
\textit{``Protect data and ensure generative AI systems and applications incorporate privacy and security by design.''} (UIUC)
\end{leftquote}

In tandem, they rely on the responsibility of individuals to adopt multiple technical safeguards for GenAI, including \textit{``encrypt[ing] sensitive data both at rest and in transit to protect it from unauthorized access,''} using \textit{``anonymized datasets''}, and \textit{``implement[ing] strict access controls, such as Multi-factor Authentication (MFA)''} (UCLA).

Yet, it is unclear how these technical safeguards are contextualized to GenAI service implementation.

\new{\textit{\textbf{Complexity of operational security in multi-department coordination.}}}
We observe universities implementing operational privacy and security measures in multiple stages of adopting and using GenAI services, which can be complicated and require multi-party involvement.
Universities such as Columbia University (and the University of Chicago) mention their implementation of security review when acquiring or procuring services that \textit{``contain functions that rely on AI to operate,''} \textit{``before they are acquired, even if the software is free.''}

Such processes can be initiated and managed by different departments including multiple IT teams within the university, e.g., Columbia IT (CUIT) or Columbia Irving Medical Center (CUIMC-IT) as mentioned in Columbia University's guidelines.

University guidelines also mention their continuous commitment to privacy and security via oversight and awareness training. 

\begin{leftquote}
\textit{``keep your training up to date, including the latest training available to you on McGill's cybersecurity awareness training platform.''} (McGill University)
\end{leftquote}

Universities often set up dedicated contact points for GenAI-related queries to address \textit{``any questions about the use of this dynamic technology.''} (Caltech) University guidelines (e.g., University of Melbourne and UIUC) may also underscore the importance of redressing GenAI harms. Nevertheless, establishing this procedure seems to be an on-going effort. 

\begin{leftquote}
    \textit{``The University of Melbourne will maintain human oversight of and responsibility for the AI tools and systems used by staff, offer appropriate avenues for redress in case of errors or misuse, and commit to procedures by which those who are impacted can access an explanation for relevant decisions.''} (University of Melbourne) 
\end{leftquote}

\subsubsection{ Guidelines shift responsibility onto individuals.}\label{ss:rq2-2}
A predominant guideline we observe across universities is the devolution of accountability to the individual (students, faculty, and staff). 

\new{\textbf{\textit{Individual accountability for making S\&P decisions.}} While some institutions continue to provide technical safeguards, individuals are also expected to be responsible for their privacy and security behaviors and decisions when interacting with GenAI tools, including not disclosing confidential information as discussed before.}
Another example is UCLA:  \textit{``It is the users' responsibility and accountability to get informed about these limitations and risks, closely review the outputs of GenAI tools...''}

To this end, universities expect individuals to choose a GenAI service that provides better privacy and security features.
However, a large number of GenAI services on the market can overwhelm individuals as they navigate alternatives.
University guidelines often mention a variety of specific GenAI service providers, including standalone GenAI chatbots, e.g., OpenAI's ChatGPT~\cite{chatgpt-url25} and Google's Gemini~\cite{deepmindGemini-url25} as well as GenAI integrated with other digital services such as CoPilot within coding IDE~\cite{githubGitHubCopilot-url25} and AI companions in online meeting apps, e.g., Zoom IQ~\cite{zoomMeetZoom-url25}.
Despite the availability of many GenAI services, a more detailed privacy and security evaluation for individual services is absent in many guidelines, except for the most popular service provider OpenAI's ChatGPT model: 

\begin{leftquote}\textit{``ChatGPT utilizes one of the largest (known) datasets to power its operation and is hence in its current version of GPT-4 seemingly capable of a variety of complex tasks''} (UM-Ann Arbor). \end{leftquote}

\new{Furthermore, users are expected to take different levels of responsibility beyond proper tool and data use, including reviewing GenAI outputs and formally citing or acknowledging GenAI use.}


\new{\textbf{\textit{Shared responsibility via licensing and risk assessment.}}}
\new{As discussed earlier, performing security assessment is a main responsibility universities are committed to, guided by regional or institutional policies like \textit{Privacy and the Need to Monitor and Access Records (SPG 601.11)} and the \textit{Institutional Data Resource Management Policy (SPG 601.12)} established by UM-Ann Arbor. As a result of such assessment and license negotiation, recommending and enforcing the use of university-licensed GenAI services seems to be a more feasible alternative that well-resourced universities offer to individuals. A few universities such as ETH Zürich, UoE, UCL, McGill, UM-Ann Arbor, Caltech have formally licensed platforms including Microsoft Copilot under enterprise agreements with Data Protection Addendums (DPAs) enabled, which shape platform behavior in alignment with institutional values such as academic integrity, data minimization, and ethical AI usage~\cite{Cohneyetal_2021}. 
An example is UM-Ann Arbor's customized U-M GPT that is \textit{``private, secure, and free for students.''} (UM-Ann Arbor) }


\new{Nevertheless, licensing can be seen as an approach by universities offloading the responsibility to GenAI vendors and transferring trust in a uniform way instead of further tailoring their own safeguards.
For instance, ETH Zürich redirects users to Microsoft's Enterprise Data Protection FAQ which reiterates, in the EU context, its commitment to GDPR~\cite{GDPR-url25}, ISO/IEC 27018~\cite{MicrosoftISO27018}, and existing Microsoft DPA~\cite{MicrosoftDPA} that is uniformly applied to all other universities. In these contexts, the main threat identified and mitigated through licensing agreements is the use of student data for training:}

\begin{leftquote}
\textit{``The freely accessible version of Microsoft Copilot offers
[...]
the ‘protected' status (with Microsoft ETH-account), whereby personal data is not used as training data,''} (ETH Zürich)
\end{leftquote}

\new{\textit{\textbf{Individuals' challenges in exercising privacy rights and options.}}}
Nevertheless, guidelines recognize the challenges of individuals exercising their privacy rights and options for privacy compliance, e.g., consent procedures, when using GenAI, reflecting universities' distrust of the GenAI service providers' privacy commitment:

\begin{leftquote}
\textit{``you are sharing your data with private companies, so you lose your control over it.''} (EPFL)
\end{leftquote}

Similarly, Columbia worries that \textit{``the technology may not respect the privacy rights of individuals...''} Universities are also concerned about individuals not accepting consent carefully:

\begin{leftquote}
\textit{``Individuals who accept click-through agreements without delegated signature authority may face personal consequences, including responsibility for compliance with terms...''} (UC Berkeley)
\end{leftquote}

McGill University explicitly recommends students to self-censor the input for GenAI \textit{``by removing personally identifying information.''}
Interestingly, we observe that UC Berkeley offers a fine-grained  classification of allowable data to be used by specific GenAI services. For example, while \textit{``Publicly-available information (Protection Level P1) can be used freely in all generative AI tools''}, \textit{``...Zoom AI Companion (pilot only) [are] approved for use with P2 information, providing everyone actively consents to its use each time.''} 
The reference also discusses disabled GenAI features which can replace human meeting attendees with bots.  
However, they fall short in discussing the possible unique privacy implications of these services for individuals to make more informed decisions, e.g., hijacked GenAI bots that produce harmful information.

Some university guidelines ask individual instructors and researchers to offer privacy rights and options when they involve GenAI into their teaching and research. 
LSE suggests instructors offer opt-out options for students.

\begin{leftquote}
\textit{``Please consider offering students a choice to opt-out of using a system if they have concerns about the cost, privacy, security or other issues related to the technology.''} (LSE)
\end{leftquote}

Likewise, UC Berkeley emphasizes the importance for researchers to obtain \textit{``explicit and informed consent from individuals before collecting, processing, or using their data in GenAI systems.''}.

\vspace{-1.5em}
\new{\subsubsection{IT departments operationalizing GenAI governance through risk assessment and monitoring.} \label{ss:rq2-3}
Our comparison between teaching and learning-focused guidelines with IT-focused guidelines highlights how university departments coordinate and operationalize GenAI governance. 
The high-level privacy and security risks, such as third-party sharing, and mitigations identified in IT guidance broadly align with those in teaching and learning-focused guidelines. While teaching and learning guidelines more often emphasize academic values and classroom integration, IT-focused guidelines translate shared privacy and security concerns into concrete assessment, deployment, and monitoring practices. }

\new{\textit{\textbf{Making privacy assessment outcome visible.}} 
Universities use IT-focused guidance to disclose tool-specific privacy and security assessment \textit{outcomes} and support safer individual adoption of specific services.
For example, the University of Oxford warns that Copilot for Microsoft 365 may increase risks to confidential university data because \textit{``access permissions are often poorly managed and this can go unnoticed''}; it provides similar assessments for tools such as Google Gemini and ChatGPT Edu.}

\new{IT departments also play a central role in managing and governing licensed GenAI services and custom GenAI systems developed by students or staff. Some host ``one-stop'' AI guidance websites that consolidate policies, approved services, and risk advice. For instance, UMich's Information and Technology Services states its role in governing and monitoring licensed and custom GenAI services, including conversational tools, API interfaces, and systems powered by contracted or open-source models for custom dataset analysis. For custom GenAI development, UofT's Information Security warns of risks from \textit{``adversary-provided training data''}, while the University of Oxford asks users to \textit{``discuss this with the Information Security GRC team in advance''} before \textit{``inputting internal University data to an external custom GPT.''}}

\new{\textit{\textbf{Inconsistent efforts in adapting existing safeguards.}}
Universities generally extend existing risk assessment processes and technical protections to GenAI rather than creating entirely new safeguards, such as applying \textit{``the same security and encryption techniques as your emails and contents of your OneDrive or Sharepoint Sites''} (UoM). However, IT departments vary in how actively they adapt and explain these frameworks for GenAI users. 16 of 19 institutions simply ask students and staff to revisit existing data protection policies, while others provide more detailed GenAI-specific interpretation. For example, Imperial College's Data Protection Officer lists privacy-sensitive activities that should be monitored and recorded under its data activity risk assessment framework, including \textit{``the versions of relevant software or firmware''} used by AI systems, the \textit{``recourse to pre-trained systems or tools provided by third parties''}, and \textit{``the cybersecurity measures put in place.''}}.

\new{\textit{\textbf{Monitoring GenAI usage.}} 
IT-focused guidelines emphasize both users' responsibility to \textit{review} GenAI outputs and institutional commitment to \textit{monitor} GenAI use that requires individual cooperation. Beyond information required for acquiring or updating GenAI services, some universities set stronger expectations for recording GenAI outputs. For example, ANU states that when GenAI outputs \textit{``meet the definition of a University record, they must be stored in the ERMS [Electronic Records Management System].''}
}

\subsection{Principles in Policy Development}
\label{ss:rq3}

\new{Considering the identified privacy and security concerns and safeguard limitations, we find that universities adopt GenAI for academic use cautiously and provisionally. Building on the privacy and security issues and safeguards discussed in \S~\ref{ss:rq1} and~\ref{ss:rq2}, this subsection examines how university, national, and international frameworks shape framing of privacy and security issues. We identify four key impact factors:}

\begin{squishenumerate}
    \item \textbf{Governance frameworks}, including privacy regulations and policies in different countries or educational associations such as university systems and groups as well as universities' relevant disciplinary procedures.
    \item \textbf{Academic values} aligned with privacy and security, such as academic integrity and other ethical values such as fairness and equality.
    \item \textbf{Risk-benefit trade-offs} in academic uses, including productivity, creativity, and personalization, when GenAI assists multiple academic uses such as teaching, learning and assessment, writing, and research.
    \item \textbf{Engaged stakeholders} in policy development, such as students, faculty, researchers, and staff in relevant departments.
\end{squishenumerate}

\new{\subsubsection{Privacy governance frameworks for GenAI vary across regions and influence guidelines}
\label{ss:rq3-1}
Universities first recognize the need for GenAI to comply with existing legal obligations. This is stated explicitly in UCT's guidance for example: \textit{``publicly available (non-personal) data may be used with AI tools. For any sensitive data, consider legal, regulatory, or contractual obligations before use''}
The awareness of the inextricability of University GenAI guidelines and existing legal frameworks also entails an acknowledgement of the need for universities to continuously update their GenAI guidelines due to the evolving technological and legal landscape:}

\begin{leftquote}
    \textit{``this document will be updated regularly and interact with other sources of policy, ethics, and governing legal authority.''} (UPenn)
    \end{leftquote}

\new{\textit{\textbf{National regulatory frameworks anchoring privacy governance rather than AI governance.}} 
At the same time, because universities' adaptation stands on existing privacy regulations, they vary across jurisdictions. 
Differences in national privacy and security regulations therefore shape the robustness and precision of university GenAI guidelines from a privacy and security standpoint.
} For universities in the US, privacy practices are governed by policies at different levels -- \textit{``Federal, state, and local laws as well as Caltech policies [that] may limit data that can be disclosed''} (Caltech). US universities frequently mention HIPAA (Health Insurance Portability and Accountability Act)~\cite{hhs-url25} which guides their sensitive data classification frameworks discussed previously in \S~\ref{ss:rq1}. US universities also rely on FERPA~\cite{ferpa2011} that protects children's education records and gives their parents the right to access and control the data. This regulation surpasses HIPAA in regulating the usage of students' medical records.

\begin{leftquote}
\textit{``It is the faculty members' responsibility to safeguard student data following all relevant regulations covered by FERPA''} (UIUC)
\end{leftquote}

\new{Beyond the US, universities similarly anchor their guidelines in national regulatory frameworks:} Singapore's PDPA is referenced by NTU~\cite{pdpa-url25}, while ANU grounds its guidance in the Australian Privacy Act (1988)~\cite{oaic-url25}. In South Africa, UCT invokes POPIA~\cite{POPIA-url25}, UK institutions reference the UK GDPR~\cite{UKGDPR-url25} alongside the Data Protection Act 2018~\cite{DPA2018-url25}, and European universities may cite the GDPR~\cite{GDPR-url25} and/or the EU AI Act~\cite{EUAIACT-url25}. In Canada, the University of Toronto adheres to the national FIPPA framework~\cite{UniversityofToronto2019}, while McGill incorporates FIPPA alongside the ``FASTER'' principles---Fair, Accountable, Secure, Transparent, Educated, Relevant---endorsed by the Chief Information Officer at the Treasury Board of Canada Secretariat~\cite{Secretariat_2024}. \new{North American universities lead policy and governance discourse, whereas European universities emphasize operational data management (\S~\ref{ss:rq2} and \S~\ref{ss:rq1}): 13 out of 18 NA universities discuss privacy regulations versus 6 out of 14 in Europe.}

\new{It is also worth noting that not all countries have regulatory frameworks tailored for GenAI usage, with a main exception in our corpus being McGill's adaptation of FIPPA according to the Canadian Government's GenAI guidance and the EU AI Act. } 
\new {Thus, attention has to be paid on the limitations of these privacy regulations, particularly in their inconsistency in regulating privacy practices. In the US, for example, universities that reference FERPA for student data protection may leave leeway that allows third-party organizations such as GenAI vendors to utilize student metadata~\cite{Willse_2023} -- a scenario that is disallowed in other legal contexts such as the Canadian FIPPA, where personal data cannot be shared to third-party organizations unless mandated by law \cite{CIGI2024}.}

\new{Despite this limitation, legal compliance can still support more systematic mitigation of the GenAI threats discussed in \S~\ref{ss:rq2}. From a cross-comparative perspective, institutions across North America, the UK, and Australia --- including UBC, MIT, McGill, UofT, and Northwestern; Oxford, ICL, and UCL; and ANU --- consistently incorporate Privacy Impact Assessments (PIAs) into their GenAI-related policies as part of their legal obligations. These assessments are explicitly communicated to the stakeholders, reflecting a more institutionalized approach to privacy governance and accountability. In contrast, publicly available guidelines from Asian universities do not overtly mention any privacy impact assessments. Instead, they may place greater responsibility on individual users to evaluate the implications of their choices. For example, at Tokyo Tech, using GenAI tools is framed as being entirely at the student's discretion.}

\textit{\textbf{Influence of cross-institutional collaboration frameworks}}
Alongside national or provincial legal frameworks, cross-university associations and collaborations with other universities also shape the process of designing guidelines. The former includes the Russell Group in the UK~\cite{russell-url25}, which has published their own set of GenAI guidelines, and US state university systems such as the University of California system, which defines \textit{``privacy and security''} among the seven core principles that guide the policy development at UC Berkeley and UCLA. Universities also make cross-references to other universities' GenAI policies to support their rules and offer complementary information. 
For example, EPFL's guideline points its reader to a FAQ page for ChatGPT policies developed by their \textit{``sister institution in Zürich.''} Similarly, Duke University directs its readers to established guidelines in other universities, noting that, in relation to best practices for GenAI attribution, \textit{``Monash University has curated a list of the various AI citation formats.''} \new{This is not a practice limited to universities in Western countries. NTU has also put together \textit{``curated examples''} from Monash University, Cornell University, KCL, and Harvard.}

\new{\textit{\textbf{Engagement with international privacy frameworks}}. In addition to inter-institutional collaboration, a number of universities also explicitly engage with international privacy frameworks. UIUC, for instance, acknowledges that relevant privacy laws span multiple jurisdictions, encompassing federal and state regulations alongside \textit{``geographic and extraterritorial international laws such as GDPR and PIPL [China's Personal Information Protection Law].''} This attention to extraterritorial reach is echoed in the policies of KAUST, Harvard and Duke, which reference GDPR and/or the EU AI Act as a relevant instrument. IC also cites the Indian Digital Act and Chinese AI Safety Framework, while Cornell mentions the EU-U.S. Data Privacy Framework. Awareness of international privacy frameworks in their GenAI guidelines highlights how universities, as globally embedded institutions, must navigate an increasingly complex, cross-jurisdictional privacy landscape.}

\subsubsection{\new {Relationship between privacy, security and other academic values}}  \label{ss:rq3-2}
Given that GenAI has profoundly impacted teaching and learning while simultaneously introducing privacy and security risks---as elaborated in \S~\ref{ss:rq1} and \S~\ref{ss:rq2}---university guidelines reflect this complexity in how they frame privacy and security in relation to other academic values.

\textit{\textbf{\new{Balancing benefits with privacy trade-offs.}}}
Universities value GenAI's positive uses and recognize that not embracing it might shortchange their students in the \textit{``AI-supported labour market and societies of tomorrow''} (ETH Zürich). However, they are also aware that GenAI tools can be a double-edged sword due to privacy and security risks. For example, the University of Chicago claims that

\begin{leftquote}
\textit{
``GenAI tools offer many capabilities and efficiencies that can greatly enhance our work...members of the university community must also consider issues related to information security, privacy, compliance, and academic integrity.''} (University of Chicago)
\end{leftquote}

\new{\textit{\textbf{Privacy and security subsumed under broader academic values.}} The tension between opportunity and risk leads universities to foreground broader academic values---particularly ethics and academic integrity---as the normative foundation for their GenAI stance.} For example, \textit{``UCL has opted to promote ethical and transparent engagement with GenAI tools rather than seek to ban them.''} 

In ETH Zürich's guidelines, privacy and security are framed not as a compliance obligation but as an expression of fairness:

\begin{leftquote}
\textit{``Fairness: Respect the privacy and copyright of the content with which you work. Refrain from disclosing copyrighted, private, or confidential information to commercial GenAI clients, unless expressly permitted.''} (ETH Zürich)
\end{leftquote}

Universities may also tie privacy and security expectations to their institutional reputation. For instance, the University of Oxford's policy declares that their \textit{``reputation stands on the trustworthiness of our research and our communications.''}

\new{\textit{\textbf{Limits of values-based framing.}}
While grounding privacy in broader values can lend it normative weight, this approach carries a risk: high-level principles are inherently open to interpretation and can sometimes substitute for concrete guidance rather than complement it. Tokyo Tech illustrates this tendency, deferring to student judgment rather than specifying expected practices:}

\begin{leftquote}
 \textit{``We expect all students to exercise good sense and judgment when using GenAI, in accordance with the Institute's spirit of “Student-Centered Learning.''} (Tokyo Tech)
\end{leftquote}

Notably, this breadth may be deliberate. Some universities explicitly design their principles to remain open-ended, as evident in the justification of the University of Melbourne's GenAI guidelines: 


 \textit{``The principles are intentionally broad: the aim is not to prescribe specific initiatives or actions, but to support decision-making across the University, and to ensure the principles can be adapted to developments in the technology. The intention is to periodically review the principles, given the ongoing evolution of AI tools.''} (University of Melbourne)

\new{\textit{\textbf{Value integration  and interaction across departments.} }
Policies from Centres for Teaching and Learning, or equivalent bodies, typically ground GenAI guidance in pedagogical values and institutional ethos. By contrast, IT or information security guidelines, often published as companion documents, foreground technical compliance and risk management.}
\new{The extent of this divergence varies across institutions. At IC, the IT-specific guidance adopts a ``privacy-first'' approach, e.g., by asking staff to minimize data collection. This emphasis on privacy is absent from its general guidance, which instead highlights values such as respect, collaboration, excellence, integrity, and innovation. In other cases such as UCLA, the guidelines from two sources show stronger integration: its IT guidance incorporates academic integrity, transparency, and ethical and equitable use, and explicitly directs users to guidance from its Center for Teaching and Learning.}

\new{Cross-referencing between IT and teaching-focused guidance also differs by region, when we look at cases we analyzed where institutions publish separate IT-focused AI guidelines, in addition to a teaching and learning one. Among Canadian universities in our corpus, all 3 institutions with IT-specific guidelines link to their teaching and learning counterparts. In the US, this practice is mixed: 5 of the 8 coded US IT-specific policies do not cross-reference. In contrast, in the UK, 6 out of 9 universities maintain dedicated ``AI hubs''---centralized landing pages that consolidate access to both teaching-focused and IT-specific guidelines under a single entry point. In Asia, KAUST provides an IT-specific GenAI guideline while 3 of the remaining 6 Asian universities address data security briefly as a principle within their teaching-centered GenAI guidance. This trend is mirrored in Australian universities in our corpus: ANU and Monash University have an IT-specific GenAI guideline, while the remaining 3 universities only cite privacy and security as one of their high-level AI usage principles.}

\subsubsection{\new{Improving guidelines with stakeholder engagement}}\label{ss:rq3-3}
Encouraging active participation from students, teaching and research staff, and other university stakeholders in both AI policy creation and its application across specific academic contexts is a pattern observed widely across university guidelines worldwide. For example, Duke University's guideline states that

\begin{leftquote}

\textit{``The Provost is consulting with faculty and staff experts on these larger questions involving generative AI systems, and welcomes debate and discussion on these issues.''} (Duke University)
    
\end{leftquote}

For these stakeholders, university guidelines acknowledge the possibility that the privacy and security considerations could be undermined by their perceived benefits, including the convenience to assist learning, education, and research. 
Therefore, we observe that university guidelines try to balance these trade-offs.
While highlighting the values for students to develop \textit{``their own skills in critical analysis, creative thinking, and rigorous research''} (University of Cambridge), university guidelines simultaneously underscore academic integrity and disciplinary procedures,  emphasizing the consequences of misusing GenAI tools.
Guidelines ask instructors to take responsibility in navigating GenAI risks for students:

\begin{leftquote}
\textit{``Instructors must educate their students about these risks, and develop plans to mitigate the negative impact of risks in their classroom if they decide to use or allow the use of GenAI in their teaching.''} (UM-Ann Arbor)
\end{leftquote}

\section{Discussion}
\label{sec:discussion}

Our research illuminates how institutions interpret privacy and security considerations and implement solutions to mitigate the risks of GenAI in academic settings, in particular the gaps in their institutional policies to safeguard GenAI. 
Our study confirms the conclusions from prior work that privacy and security are a key consideration when adopting GenAI in academic settings~\cite{Jin_Yan2025, McDonald_Johri_Ali_Collier_2025, ali2025-cse25, jiao-nature25}.
Moreover, our work contributes novel findings compared to prior studies that treated privacy and security concerns as epiphenomenal to broader pedagogical and implementation discussions~\cite{Wang_Dang_Wu_Mac_2024}, and a gap in these studies is the distinct lack of granularity in their analysis framework for privacy and security \cite{McDonald_Johri_Ali_Collier_2025, Jin_Yan2025}. 
As a result, prior findings may superficially identify university awareness of the topic about privacy and security but does not analyze the extent of their understanding, assessment and the challenges faced in mitigating privacy and security issues. Recommendations offered thus also remain at the procedural level, such as how to clarify confidential data types for ensuring greater compliance, without delving into the underlying technical infrastructures and inadequacy of existing privacy frameworks in adapting to GenAI capacities. 

Furthermore, there is a dearth of privacy and security research that applies specifically to emerging GenAI services. 
The GenAI use in  universities presents a distinct context of privacy and security compared to enterprise settings~\cite{schneider2020ai, taeihagh2025governance}: Although both share concerns such as IP leakage, universities stress long-term learning impacts and student privacy harms, whereas enterprises prioritize efficiency and customer satisfaction. Companies also rely on universities for early ethics training while maintaining clearer models for GenAI-related cyber-attacks.
Compared to traditional education technologies such as exam proctoring systems~\cite{chanenson2023uncovering, shioji-sp24,kumar2019edtechelementary}, the adoption and usage of GenAI is much more flexible and dynamic to accommodate individual requirements and varying application contexts. GenAI amplifies the concerns flagged in the former, such as privacy intrusion and monitoring~\cite{Yang_Hasan_2023}, and the university policies aimed at maintaining control over vendors of education technologies~\cite{kelso2024trust} can be unfeasible in this new context.

\subsection{Recommendations}
\label{ss:recos}

\noindent \textbf{Improving privacy and security assessment frameworks for GenAI in academic settings.}
We found that universities often assess GenAI risks through existing frameworks (e.g., FERPA) and their internal data classification and audit procedures \new{under the influences of regional and international privacy regulations (\S~\ref{ss:rq1} and \S~\ref{ss:rq3}). 
Regulatory development varies across regions: the US has been more advanced in educational privacy governance, while the EU has taken the lead in AI governance. Interestingly, national privacy and GenAI regulatory developments influence higher education across political borders, both among allies such as Canada and the EU, and between strategic competitors such as the US and China, as reflected in UIUC's recognition of China's PIPL (\S~\ref{ss:rq3-1}).
Although it takes time for GenAI regulations to mature, we argue that these influences can extend beyond shaping how university guidelines frame privacy and security risks, supporting the systematic adoption and improvement of local frameworks, such as student- and staff-facing privacy assessment toolkits, to reduce inconsistencies in regulating privacy practices (\S~\ref{ss:rq3-1} and~\ref{ss:rq3-3}).
}
\new{However, our results further show that  legal and institutional instruments may fall short of addressing the emerging privacy and security challenges introduced by GenAI -- particularly its enhanced capacity to infer sensitive information from minimal or aggregated data~\cite{staab2023beyond}. 
}
Such capabilities exacerbate risks of re-identification, discrimination, and misuse, which are highlighted in universities' GenAI guidelines (c.f. \S~\ref{ss:rq1}). 
For example, GenAI could infer students' demographics and sensitive information (gender, race, and health)~\cite{lee-chi24} indirectly from their linguistic patterns revealed in dialogues without accessing student records.  
Effective governance requires expanding risk models to reflect these GenAI-specific threats and adopting a multi-stakeholder approach, engaging technical teams, legal experts, and academic departments in policy design.
More than performing security review during procurement and adoption (as currently highlighted by the guidelines), universities should develop and improve their oversight process to identify new privacy and security risks and remediate harms with fast responses (\S~\ref{ss:rq2-3}). 
Instead of relying only on prolonged human review, universities can leverage licensed GenAI services to detect anomalous model interactions in a privacy-preserving manner~\cite{zhang-tsem25}.

\vspace{1mm}
\noindent \textbf{Tailoring technical safeguards to academic contexts.}
\new{Most institutions still rely on traditional IT infrastructure (e.g., account authorization and access control to disable GenAI features), which they have deployed for conventional educational technologies such as LMS, to primarily mitigate risks about data access and breaches. 
However, this approach does not adequately address the evolving threat landscape posed by GenAI and enforce compliance off-campus (\S~\ref{ss:rq2}),  as GenAI systems synthesize data, introducing uncertainties and new or amplified privacy risks such as identification and physiognomy-related harms~\cite{lee-chi24}. As such, monitoring GenAI outputs becomes a new lever to govern GenAI risks. 
However, universities are also not confident in the reliability of novel safeguards such as AI detectors that ease the efforts of output monitoring.}
Furthermore, \S~\ref{ss:rq3-2} discusses the specific requirements of the universities to align GenAI models with academic values and expect privacy-by-design~\cite{gurses-cpdp11} for these models, which currently falls short in GenAI development.
We notice several opportunities to address these requirements given recent technology advances for GenAI safeguards.
\new{First, while not perfect, AI detectors could improve their reliability in academic contexts by fine-tuning on expert feedback provided during assessment to improve decision-making with human in the loop~\cite{dugan-aaai23}.
Second, model watermarking, which actively embeds a robust signal invisible to humans into its generated content
for verification~\cite{kuditipudi2023robust}, can be leveraged to ensure compliant use of licensed models even off-campus.
Third, differentially private AI model alignment~\cite{shen-arxiv23,wu2023privately} can be applied during model training to ensure privacy and security-by-design along with other academic values tailored to university needs.
Improved alignment strategies potentially offer users an efficient way to mitigate risks.
Recent commercial approaches such as Google's SynthID~\cite{dathathri2024scalable} and VaultGemma~\cite{sinha2025vaultgemma} shed light on the practicality of these directions.}

\vspace{1mm}
\noindent \textbf{Reducing infrastructure disparities through strategic partnerships for integrating GenAI safeguards.}
Developing and implementing effective GenAI safeguards often requires significant technical and human resources, which many smaller institutions may lack~\cite{Wang_Dang_Wu_Mac_2024}. 
Nevertheless, we observe good examples of universities and regulators encouraging  partnerships in their policy creation, e.g., McGill University's adoption of Canada's FASTER framework illustrates the value of national policy alignment (\S~\ref{ss:rq3-1}).

We argue that these collaborations and partnerships can be deepened to benefit less-resourced institutions by facilitating knowledge sharing and joint infrastructure development, thus securing their GenAI adoption, e.g., via sharing of GenAI services co-hosted and overseen by  universities.
The current collaborative partnerships are developed mostly within existing systems or regional consortia.
Broader academic networks, like the Russell Group in the UK, could serve as operational hubs to support less-resourced institutions~\cite{huggins-erd12}.
\new{Nevertheless, we identify several open challenges to solve when supporting less-resourced institutions. First, beyond resources disparities, we argue that the differing academic and educational missions may influence their positions and focus on privacy and security. For example, compared to research-active universities, the privacy concerns of administrative data such as student records might be more prevalent in professional schools than the IP concerns of GenAI handling research datasets (\S~\ref{ss:rq1-1}); second, the concern of lacking privacy and security awareness and literacy for staff and students, who take major responsibilities to avoid GenAI misuse, may be further amplified in less-resourced institutions~\cite{rezgui2008information};  furthermore, additional privacy and compliance issues need to be considered in these partnerships, e.g., data sharing across universities and regions (\S~\ref{ss:rq3-1}).}

\vspace{1mm}
{\noindent \textbf{Improving privacy and security communications for GenAI in higher education through analogies and examples.}
Guidelines highlight the importance of individual accountability and awareness for privacy and security of GenAI adoption, as enforcing central rules that govern every specific GenAI application seems unrealistic and non-ideal (\S~\ref{ss:rq2}).
As such, effective privacy and security communications become crucial for students and instructors to contextualize their GenAI application, e.g., at the course level.

However, disparities in student and instructor preparedness---particularly between technical and non-technical disciplines~\cite{ ali2025-cse25, elrayah-ijcc23}---undermine consistent privacy and security awareness and enforcement.
\new{Our additional analysis shows that guidelines we analyzed require college-level comprehension, with regional variations: lower average Flesch reading ease scores~\cite{farr1951simplification} (more difficult) are found for North American guidelines (27.2) than those from Europe (38.3) and Asia (33.7) in Table~\ref{tab:policies}.}
Though privacy and security policies can be complicated to understand in a GenAI guideline due to its technical nature, our findings in \S~\ref{ss:rq1-2} show that universities use analogies to help stakeholders make sense of emerging GenAI technologies, which can be further adapted for privacy and security communication. 
For example, NUS and UofT effectively communicate the level of appropriate use of GenAI services in the teaching and learning context by drawing analogies between using GenAI and \textit{``using calculators''} or \textit{``consulting peers.''}
This approach situates the new technology in \textit{``a familiar historical and socio-technical context,''} understood as a linear development of pedagogical methods~\cite{Schwarz-Plaschg_2018}.

In the same manner, analogies can be developed to improve the accessibility of otherwise abstract and technical jargon-laden data privacy and security measures to support students, staff and researchers in making optimal privacy and security protection decisions~\cite{chen-chi24}. For example, guidelines could help people understand how their personal data can be \textit{``unlearned''} by GenAI models as a new approach to exercise their privacy rights~\cite{bourtoule-sp21}.
In addition, examples such as template syllabi (e.g., from UM-Ann Arbor) and sensitive data types are observed as another starting point for university guidelines to clarify complex privacy and security concepts (\S~\ref{ss:rq1} and \S~\ref{ss:rq2}).
Continuous professional development remains crucial to the cause: universities should provide flexible communications and training, including micro-training for data privacy embedded in the interaction with university-licensed GenAI~\cite{shamir-pde22, chaipidech-ceai22}.
Example user prompts can also be provided to improve privacy and security alongside productivity in GenAI interaction~\cite{zamfirescu-chi23}.

\subsection{Future Work}
\label{ss:future}

\vspace{1mm}
\noindent \textbf{Longitudinal and multi-level policy measurement and analysis.}
A promising direction for future research involves tracking how GenAI governance evolves across time and institutional levels. Longitudinal studies, enabled by automated policy collection and natural language processing (NLP) pipelines, could uncover how universities adapt GenAI-related guidelines in response to shifts in law, pedagogy,  technology, as well as privacy and security incidents of GenAI. Multi-level comparisons, across institutional policies, departmental guidelines, course syllabi, and broader privacy frameworks like FERPA and GDPR~\cite{ferpa2011,GDPR-url25}, can reveal alignment gaps and structural inconsistencies in GenAI governance. Another promising avenue for research would be to examine where such discrepancies stem from, for example, tensions between regulatory levels and varying national attitudes towards restrictive regulations. A further exploration of trade-offs across different contexts or overarching governance frameworks would provide valuable insights.

\vspace{1mm}
\noindent \textbf{Understanding stakeholder perceptions on policy-making and enforcement of GenAI privacy and security.}
Future work should further study how different stakeholders (students, instructors, administrators, and service providers) understand and engage with universities' GenAI policies and enforcement, e.g., through interviews and surveys. 
Identifying gaps due to misconceptions and conflicting interests about privacy and security risks can inform the design of more accessible, targeted communication strategies, as well as business and service models to improve licensing. 

\vspace{1mm}
\noindent \textbf{Designing and evaluating privacy-preserving GenAI systems for education in a human-centric way.}
Technically, a key research challenge lies in building usable GenAI systems and safeguards that are aligned with academic values and personalized to user needs. 
Future research can involve students and educators in different phases, from co-designing prototypes to end-to-end system evaluation, when developing privacy and security measures for GenAI, e.g., for managing data sharing, interaction logs, and model behavior transparently.
Balancing privacy with personalization and creative affordances is also central to making GenAI both effective and responsible in educational contexts.

\section{Conclusions}

GenAI has become increasingly popular in higher education. 
It enables new education paradigms and boosts people's efficiency, productivity, and creativity in various educational activities.
However, there are significant privacy and security risks accompanying the wide adoption of GenAI.
To understand how universities address these emerging privacy and security risks and evaluate mitigation strategies, we performed an in-depth qualitative analysis of GenAI guidelines from  43 universities around the world. 
Our study uncovered novel insights about the gaps universities face in safeguarding GenAI usage, including the limitations in existing technical and procedural measures to mitigate risks and the alignment of privacy and security with other academic values in institutional policy development. 
These findings further motivated recommendations to improve assessment frameworks, design tailored technical solutions for academic contexts, and enhance stakeholder communication and awareness for GenAI privacy and security in academic settings.

\begin{acks}
We would like to acknowledge the support from Madelyn Rose Sanfilippo, and our anonymous reviewers and editor. Bei Yi Ng, Jiarui Li, Xinyuan Tong, and Jingjie Li would like to thank the Generative AI Laboratory, University of Edinburgh. For Kevin Ye, Gauthami Yenne, and Varun Chandrasekaran, this research received no specific grant from any funding agency in the public, commercial, or not-for-profit sectors.
The authors used generative AI-based tools to revise the text, improve flow and correct grammatical errors.
\end{acks}

\bibliographystyle{plain}
\bibliography{main}

\appendix

\clearpage
\onecolumn
\section{Appendix}
\subsection{Complementary Sample}
\label{ss:appendix-1}

{
\scriptsize
\begin{center}

\begin{tabular}{p{1.8cm} p{2cm} p{7cm} p{5.2cm}}
\toprule
\textbf{Continent} & \textbf{Country/Region} & \textbf{University} & \textbf{Document Title} \\
\midrule
Africa & South Africa & University of Cape Town (UCT) & \href{https://icts.uct.ac.za/ai-guidelines-support-admin-staff}{ICTS AI Guidelines} \\
Asia & Saudi Arabia & King Abdullah University of Science and Technology (KAUST) & \href{https://it.kaust.edu.sa/initiatives/third-party-application-security}{New Security Controls on Third-Party App Plugins \& AI Bots} \\
Europe & United Kingdom & Imperial College London (IC) & \href{https://www.imperial.ac.uk/media/imperial-college/administration-and-support-services/legal-services-office/public/data-protection/DPA-CoP-08---Use-of-Artificial-Intelligence-%28AI%29.pdf}{DPO: Considerations for the Use of AI} \\
Europe & United Kingdom & London School of Economics (LSE) & \href{https://info.lse.ac.uk/staff/services/Policies-and-procedures/Assets/Documents/aiLegRegGui.pdf}{DTS: AI Legal and Regulatory Guidance } \\
Europe & United Kingdom & University of Edinburgh (UoE) & \href{https://information-services.ed.ac.uk/help-consultancy/is-skills/digital-safety-wellbeing-and-citizenship/ai-safety/ai-privacy}{IS AI Safety: Data Privacy} \\
Europe & United Kingdom & University College London (UCL) & \href{https://www.ucl.ac.uk/information-security/information-security-policy/guidance-use-third-party-ai-services-ucl}{IS Guidance on Use of Third Party AI Services} \\
Europe & United Kingdom & University of Oxford  & \href{https://www.infosec.ox.ac.uk/use-generative-ai-services-such-as-chatgpt-safely}{IS Use Generative AI Services} \\
North America & Canada & McGill University  & \href{https://www.mcgill.ca/it/ai/ai-guidelines}{IT Service AI Guidelines
}\\
North America & Canada & University of British Columbia (UBC) & \href{https://genai.ubc.ca/guidance/principles/}{Principles for the use of GenAI tools} \\
North America & Canada & University of Toronto (UofT) & \href{https://security.utoronto.ca/governance/guidelines/use-ai-intelligently/}{IS Use Artificial Intelligence Intelligently} \\
North America & United States & California Institute of Technology (Caltech)& \href{https://www.CalTech.edu/campus-life-events/campus-announcements/guidance-on-the-use-of-generative-ai-and-large-language-model-tools}{Guidance on the Use of GenAI \& LLM Tools} \\
North America & United States & Carnegie Mellon University (CMU)  & \href{https://www.cmu.edu/computing/services/ai/meet-ai/secure-ai.html}{CIO Use AI Safely at CMU} \\
North America & United States & Cornell University  & \href{https://it.cornell.edu/ai/ai-guidelines}{IT Guidelines for Artificial Intelligence} \\
North America & United States & Duke University  & \href{https://oit.duke.edu/services-tools/artificial-intelligence/security-and-privacy/}{OIT Artificial Intelligence Security and Privacy} \\
North America & United States & University of California, Los Angeles (UCLA) & \href{https://dts.ucla.edu/initiatives/ai/ai-use-recommendation-guide}{DTS AI Use and Recommendations} \\
North America & United States & University of Illinois Urbana-Champaign (UIUC)   & \href{https://www.cybersecurity.illinois.edu/privacy-considerations-for-generative-ai/}{CIO Privacy considerations for Generative AI} \\
North America & United States & University of Michigan-Ann Arbor (UM-Ann Arbor) & \href{https://its.umich.edu/computing/ai}{ITS AI Guidelines} \\
North America & United States & University of Michigan-Ann Arbor (UM-Ann Arbor) & \href{https://safecomputing.umich.edu/protect-the-u/safely-use-sensitive-data/AI-and-UM-Data}{ITS Artificial Intelligence and U-M Data} \\
Oceania & Australia & Monash University  & \href{https://www.monash.edu/cybersecurity/awareness-and-training/data/use-ai-responsibly}{Cybersecurity: Use AI Responsibly} \\
Oceania & Australia & Australian National University (ANU) & \href{https://www.anu.edu.au/privacy/training-and-resources/generative-ai-and-data-governance}{CGRO Generative AI and Data Governance} \\
\bottomrule
\end{tabular}
\captionof{table}{\new{Additional GenAI guidelines contributed by IT departments.}}
\label{tab:it_policies}
\end{center}
}

\end{document}